\documentclass[aps,prl,reprint,superscriptaddress,nofootinbib]{revtex4-2}

\usepackage{amsmath,amssymb,bm}
\usepackage{graphicx}
\usepackage{dcolumn}
\usepackage[colorlinks=true,
linkcolor=blue,
citecolor=blue,
urlcolor=blue]{hyperref}
\usepackage{siunitx}
\usepackage{xcolor}

\begin{document}
	
	\title{Predominant Nuclear Excitation by Electron Capture Driven by Beam-Induced Return Currents}
	
	\author{Y. Y. Xu}
	\thanks{These authors contributed equally to this work.}
	\affiliation{College of Science, National University of Defense Technology,
		Changsha 410073, China}
	\affiliation{Shenzhen Key Laboratory of Ultraintense Laser and Advanced Material Technology,
		Center for Intense Laser Application Technology, and College of Engineering Physics,
		Shenzhen Technology University, Shenzhen 518118, China}
	
	\author{J. T. Qi}
	\thanks{These authors contributed equally to this work.}
	\affiliation{Shenzhen Key Laboratory of Ultraintense Laser and Advanced Material Technology,
		Center for Intense Laser Application Technology, and College of Engineering Physics,
		Shenzhen Technology University, Shenzhen 518118, China}
	
	\author{H. Peng}
	\affiliation{Shenzhen Key Laboratory of Ultraintense Laser and Advanced Material Technology,
		Center for Intense Laser Application Technology, and College of Engineering Physics,
		Shenzhen Technology University, Shenzhen 518118, China}
	
	\author{K. Jiang}
	\affiliation{Shenzhen Key Laboratory of Ultraintense Laser and Advanced Material Technology,
		Center for Intense Laser Application Technology, and College of Engineering Physics,
		Shenzhen Technology University, Shenzhen 518118, China}
	
	\author{R. Li}
	\affiliation{Shenzhen Key Laboratory of Ultraintense Laser and Advanced Material Technology,
		Center for Intense Laser Application Technology, and College of Engineering Physics,
		Shenzhen Technology University, Shenzhen 518118, China}
	
	\author{Q. Xiao}
	\affiliation{College of Science, National University of Defense Technology,
		Changsha 410073, China}
	
	\author{T. P. Yu}
	\email{tongpu@nudt.edu.cn}
	\affiliation{College of Science, National University of Defense Technology,
		Changsha 410073, China}
	
	\author{T. W. Huang}
	\email{taiwu.huang@sztu.edu.cn}
	\affiliation{Shenzhen Key Laboratory of Ultraintense Laser and Advanced Material Technology,
		Center for Intense Laser Application Technology, and College of Engineering Physics,
		Shenzhen Technology University, Shenzhen 518118, China}
	
	\begin{abstract}
		Predicted nearly five decades ago, nuclear excitation by electron capture (NEEC) remains experimentally elusive because its weak resonant signal is obscured by competing electron-driven excitation channels. Here we show that beam-induced surface return currents naturally overcome this limitation by creating a self-organized resonant electron source for the $8.36$-eV nuclear transition in solid-density $^{229}\mathrm{Th}$. A relativistic electron beam drives localized return currents along the solid surface, which simultaneously generate capture vacancies through impact ionization and provide resonant electrons for NEEC via a drifted Fermi distribution. Their spatial separation from the driving beam and Pauli exclusion principle strongly suppress competing nuclear excitation by inelastic electron scattering. For experimentally available parameters, we predict $2.40\times10^{5}$ NEEC events with a NEEC fraction of $97.68\%$, and show that the excitation yield can be tuned over several orders of magnitude with preserving NEEC dominance. These results establish beam-induced return currents as a controllable route to resonant nuclear excitation in solids and open a new avenue for studying electron-driven nuclear processes.
	\end{abstract}
	
	\maketitle
	
	Controlling nuclear excitation through electronic degrees of freedom provides a powerful means to manipulate nuclear states and explore electron--nucleus interactions \cite{Palffy01112010,PhysRevC.90.015802,Yang2025}. Among the proposed mechanisms, nuclear excitation by electron capture (NEEC), the time-reversed process of internal conversion, is particularly attractive because it resonantly couples free electrons to nuclear transitions through atomic capture \cite{GOLDANSKII1976393,PhysRevA.73.012715,PhysRevLett.99.172502}, and provides a unique opportunity to bridge atomic and nuclear physics. NEEC is especially promising for low-energy nuclear transitions, such as the $8.36$-eV isomer in $^{229}\mathrm{Th}$ \cite{PhysRevLett.132.182501,Zhang2024Nature,10.3389/fphy.2023.1166566,Xu2026}, which has stimulated intense interest for applications in nuclear clocks, quantum control, and precision tests of fundamental physics \cite{E.Peik_2003,PhysRevLett.108.120802,Peik_2021,PhysRevLett.97.092502,Hiraki2024}. 
	
	Despite nearly five decades of theoretical and experimental efforts, an unambiguous observation of NEEC remains elusive \cite{Chiara2018Nature,PhysRevLett.122.212501,PhysRevLett.128.242502,kbf5-6fcl} because its weak resonant signature is difficult to isolate from competing excitation channels \cite{PhysRevC.79.014604,PhysRevC.106.044604,PhysRevLett.124.242501,PhysRevC.110.064621,PhysRevLett.112.082501,10.1063/1.4935294}. Existing approaches have sought to generate the resonant electrons and capture vacancies required for NEEC in a variety of physical environments, including beam-target interactions \cite{PhysRevC.95.034312,PhysRevLett.127.042501}, laser-produced plasmas \cite{PhysRevLett.120.052504,PhysRevE.97.063205,PhysRevLett.130.112501}, electron-beam ion traps \cite{Wang2023EBIT,r9tv-yb8h}, and storage rings \cite{Yang2024StorageRing,PhysRevC.110.014330}. These platforms have demonstrated complementary advantages by exploiting evolving charge states, tunable electron energies, or repeated electron-ion interactions. However, none has yet achieved both a high NEEC yield and a dominant NEEC contribution under experimentally relevant conditions. Realizing experimentally observable NEEC therefore requires a controllable electron source that simultaneously provides a high flux of resonant electrons, efficiently generates capture vacancies, and intrinsically suppresses competing excitation channels.
	
	\begin{figure}[h]
		\centering
		\includegraphics[width=0.89\columnwidth]{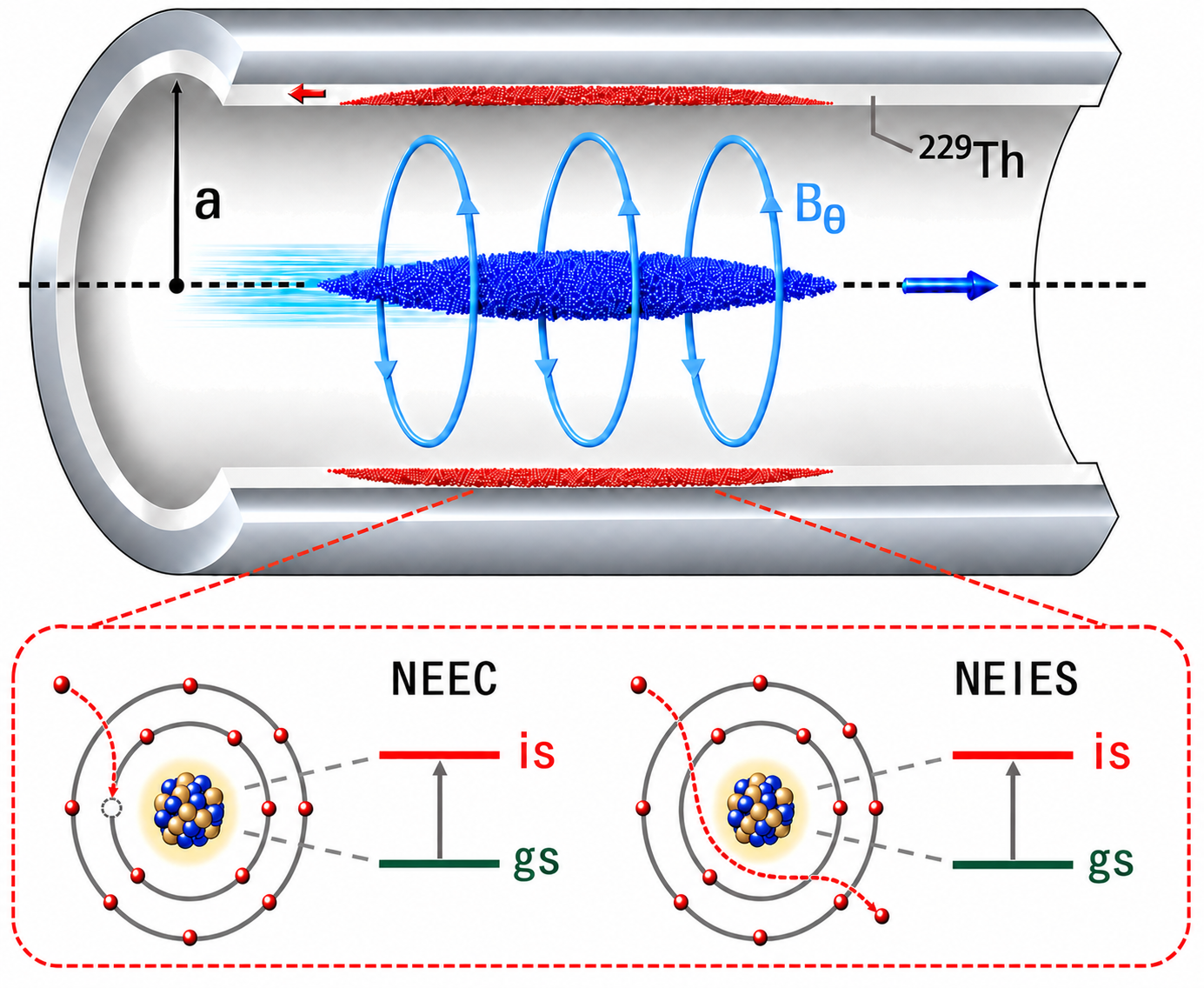}
		\caption{
			Concept of NEEC driven by beam-induced surface return currents.
			A relativistic electron beam propagating through a hollow channel
			drives a localized return current in the solid-density
			$^{229}\mathrm{Th}$ layer lining the inner surface. The
			return-current electrons create capture vacancies by impact
			ionization and supply electrons at the NEEC resonance energies,
			while also driving the competing NEIES channel.
		}
		\label{fig:concept}
	\end{figure}
	
	In this Letter, we identify beam-induced surface return currents as a self-organized resonant electron source for high-yield and dominant NEEC. As illustrated in Fig.~\ref{fig:concept}, a relativistic electron beam (REB) propagates through a hollow channel coated with a solid-density $^{229}\mathrm{Th}$ layer. The REB drives a localized return current along the thorium surface, where the conduction electrons simultaneously generate capture vacancies through impact ionization and provide a high flux of resonant electrons via a drifted Fermi distribution. Because the driving REB remains spatially separated from the thorium layer, direct beam-induced nuclear excitation is strongly suppressed, leaving nuclear excitation by inelastic electron scattering (NEIES) as the main competing channel. Furthermore, the low-energy return-current electrons obey Fermi-Dirac statistics, so that Pauli blocking substantially reduces the available final states for NEIES while leaving NEEC largely unaffected. We develop a self-consistent model of the coupled return-current and vacancy dynamics, and show that this mechanism simultaneously enhances the NEEC yield and suppresses competing excitation, establishing a broad parameter regime in which NEEC becomes the dominant electron-driven nuclear excitation channel.
	
	\textit{Return-current electron source.---}
	To determine the electron source responsible for NEEC, we first model the beam-driven return current generated at the inner surface of the hollow target. We employ axisymmetric cylindrical coordinates $(x,r)$, where $x$ is the beam-propagation direction, $r$ the radial coordinate, and $a$ the channel radius. The beam profile is described in the comoving coordinate $\xi=t-x/c$. For a Gaussian REB with total charge $Q$, rms duration $\tau_x$, and rms radius $\sigma_r$, the line-charge density is $\lambda(\xi)=\frac{Q}{\sqrt{2\pi}\,c\tau_x}\exp\!\left(-\frac{\xi^2}{2\tau_x^2}\right)$, corresponding to a beam current $I_b(\xi)=c\lambda(\xi)$, with peak value $I_{b,0}=Q/(\sqrt{2\pi}\tau_x)=2\pi e c n_b\sigma_r^2$ \cite{StupakovPenn2018}. 
	
	When the REB propagates into the hollow target, its magnetic field will induce an axial screening current (or return current) in the conducting thorium layer, whereas its radial electric field is compensated primarily by surface charges \cite{PhysRevLett.106.105002,PhysRevLett.107.135005,Si23}. For a channel radius much larger than the beam radius, the incident fields at the wall are well approximated by their vacuum expressions, $E_r^{\mathrm{vac}}(a,\xi)\simeq \lambda(\xi)/(2\pi\varepsilon_0a)$ and $B_\theta^{\mathrm{vac}}(a,\xi)\simeq E_r^{\mathrm{vac}}(a,\xi)/c$. As the return current decays exponentially into the target over the electron skin depth $\delta=c/\omega_p$, the return-current density is then obtained as
	\begin{equation}
		J_x(r,\xi)=
		\frac{\omega_p Q}{(2\pi)^{3/2}c\tau_xa}
		\exp\!\left(-\frac{\xi^2}{2\tau_x^2}\right)
		\exp\!\left[-\frac{r-a}{\delta}\right],
		\label{eq:return_current}
	\end{equation}
	where $r-a$ denotes the depth below the inner surface. The detailed derivation is presented in the Supplemental Material \cite{SupplementalMaterial}. \nocite{STEWART19661203,Ammosov1986ADK,
		PhysRevA.59.569,PhysRevA.98.043407} For nonuniform electron densities, the radial attenuation can be evaluated numerically using the local skin depth. 
	
	Equation (\ref{eq:return_current}) shows that the return current is confined to a skin-depth layer and directly scales with the instantaneous beam current. This localized current layer provides both the electron density and the spatial localization required for efficient NEEC. The return-current density also defines the local drift motion of the conduction electrons. For a local electron density $n_e$, the drift speed is $u_d=J_x/(e n_e)$, with the electron drift opposite to the return-current direction. The corresponding nonrelativistic drift-energy, $E_d=m_eu_d^2/2$, characterizes the shift of the electron distribution and determines its overlap with the NEEC resonance, where $m_e$ refers to the rest electron mass. 
	
	To benchmark the return-current model, we perform kinetic simulations using the Fourier-Bessel particle-in-cell code FBPIC \cite{LEHE201666}. Its quasi-cylindrical spectral formulation is well suited to the present beam-channel geometry and evolves the particles and electromagnetic fields self-consistently, which thus can reproduce the beam-driven return current without prescribing its profile. In the simulations, a moving window of length $24~\mu\mathrm{m}$ and radial extent $7.3~\mu\mathrm{m}$ was resolved with spatial steps $\Delta x=2~\mathrm{nm}$ and
	$\Delta r=1~\mathrm{nm}$, using a time step $\Delta t=0.9\Delta x/c$. Open boundary conditions were applied in both the longitudinal and radial directions. The channel radius is set as $a=7~\mu\mathrm{m}$ and the target wall occupied $7.0\le r<7.3~\mu\mathrm{m}$ and consisted of a charge-neutral plasma with $\mathrm{Th}^{4+}$ ions of density $n_i=3.04\times10^{28}~\mathrm{m^{-3}}$ and electrons of density $n_{e0}=4n_i$. Each species was represented by 32 macroparticles per cell. The Gaussian electron beam employed the parameters of $\sigma_x=3~\mu\mathrm{m}$, $\sigma_r=2~\mu\mathrm{m}$, $n_b=2.23\times10^{19}~\mathrm{cm^{-3}}$, and the Lorentz factor $\gamma_b=500$ (approximately $255~\mathrm{MeV}$), and was represented by $2\times10^6$ macroparticles. Ionization and collisions were omitted in these simulations so that they benchmark only the collective electromagnetic response, while ionization and vacancy dynamics are incorporated subsequently in the NEEC calculation.
	
	\begin{figure}[h]
		\centering
		\includegraphics[width=\columnwidth]{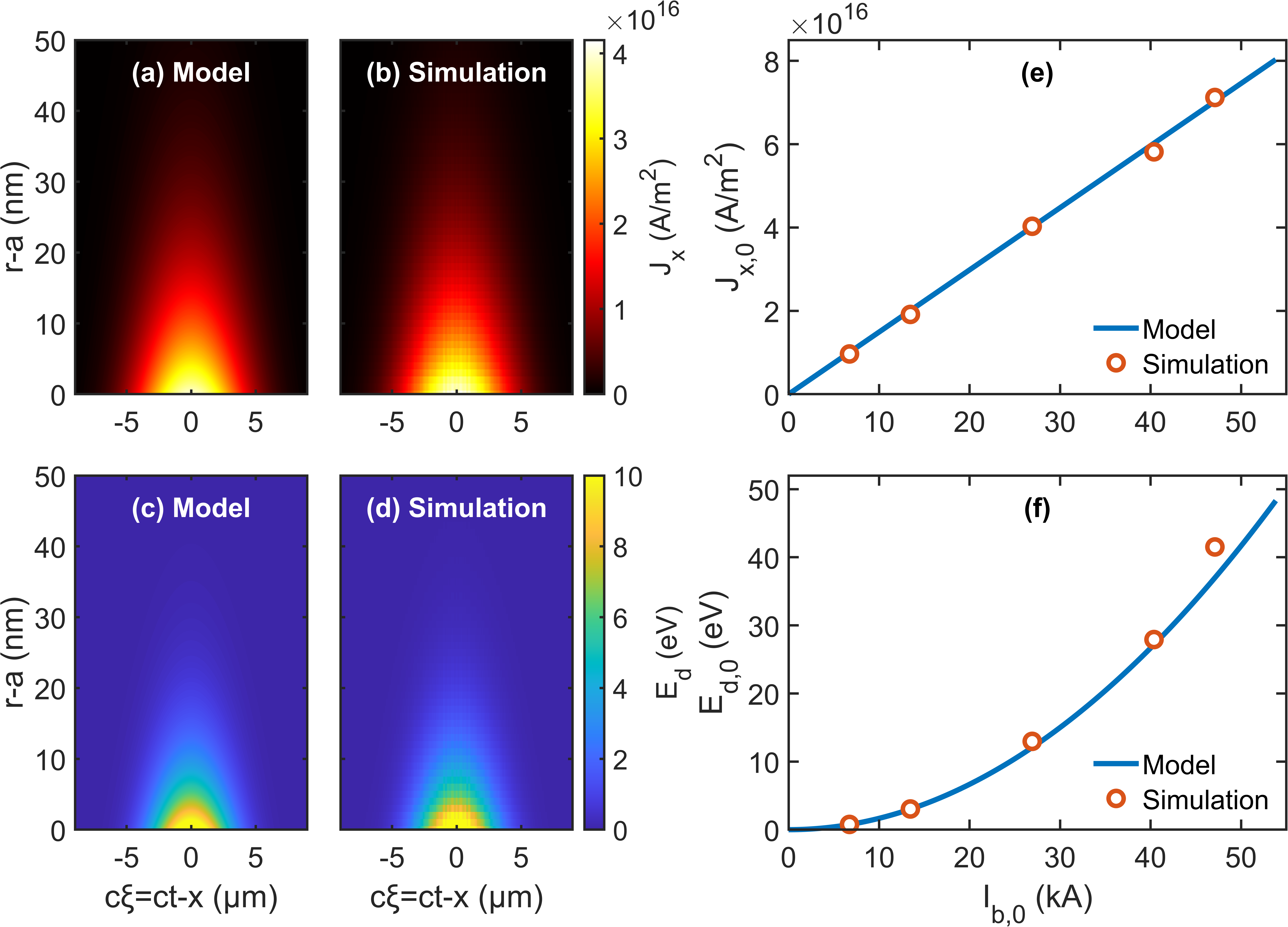}
		\caption{
			Benchmark of the reduced return-current model against FBPIC
			simulations.
			(a),(b) Near-wall return-current density from the reduced model
			and FBPIC, respectively.
			(c) Drift-energy distribution predicted by the reduced model.
			(d) Electron kinetic-energy distribution obtained from the FBPIC
			particle diagnostic.
			(e),(f) Peak inner-surface return-current density $J_{x,0}$ and
			drift energy $E_{d,0}$ as functions of the peak beam current
			$I_{b,0}$.
			Lines denote the reduced-model predictions, and symbols denote
			the FBPIC results. In (e) and (f), $I_{b,0}$ is varied by changing $n_b$, with all other parameters held fixed.
		}
		\label{fig:return_validation}
	\end{figure}
	
	Figure~\ref{fig:return_validation} compares the theoretical model of Eq.~(\ref{eq:return_current}) with FBPIC simulations. The model accurately reproduces the spatial structure of the beam-driven return current and the corresponding drift-energy distribution of the conduction electrons, as shown in Figs.~\ref{fig:return_validation}(a)-(d). Over a broad range of peak beam currents, Figs.~\ref{fig:return_validation}(e) and (f) indicate that the predicted peak return-current density and drift energy remain in excellent agreement with the simulations, following the expected scalings $J_{x,0}\propto I_{b,0}$ and $E_{d,0}\propto I_{b,0}^{2}$. Such quantitative agreement validates the theoretical model on collective surface response and provides a reliable description of the return-current electron source for the NEEC calculations presented below.
	
	\textit{Resonant excitation in solid-density thorium.---}
	To describe the electron population responsible for NEEC, we model the conduction electrons locally as a zero-temperature degenerate Fermi gas. The beam-driven return current shifts the occupied Fermi sphere by the drift momentum $m_e\mathbf{u}_d$ \cite{AshcroftMermin1976,Zebarjadi2007,PhysRevB.81.125302}, leading to
	\begin{equation}
		f_d(\mathbf{p})=
		\Theta\!\left[
		p_F-\left|\mathbf{p}-m_e\mathbf{u}_d\right|
		\right],
		\label{eq:drifted_fermi}
	\end{equation}
	where $\mathbf{p}$ is the electron momentum, $\Theta$ is the Heaviside step function, and $p_F=\hbar(3\pi^2n_e)^{1/3}$ is the local Fermi momentum. Projection onto the kinetic energy $E=p^2/(2m_e)$ gives the normalized energy distribution $f_E(E)$ with $\int_0^\infty f_E(E)\,dE=1$. The corresponding Fermi energy and velocity are $E_F=p_F^2/(2m_e)$ and $v_F=p_F/m_e$. In this case, $f_E(E)$ determines both the resonant electron population available for NEEC and the Pauli blocking of competing electron-scattering processes.
	
	At the solid density of thorium, $\rho_0=11.7~\mathrm{g\,cm^{-3}}$, the finite-density electronic structure is calculated with the average-atom code atoMEC \cite{PhysRevResearch.4.023055,callow2022}, yielding approximately four conduction electrons per atom together with the localized capture states. Vacancy production and charge-state evolution are described by a sequential rate model based on modified binary-encounter Bethe (MBEB) cross sections \cite{GUERRA20121}. The resulting evolution of the conduction-electron density updates the local Fermi momentum and skin depth, thereby modifying the return-current profile and the drifted electron distribution self-consistently. Further details are provided in the Supplemental Material \cite{SupplementalMaterial}. Here the field ionization is negligible because the radial beam field is screened by the surface response, while the residual longitudinal field remains too weak to produce appreciable ionization during the beam transit \cite{SupplementalMaterial}.
	
\begin{figure}[h]
		\centering
		\includegraphics[width=\columnwidth]
		{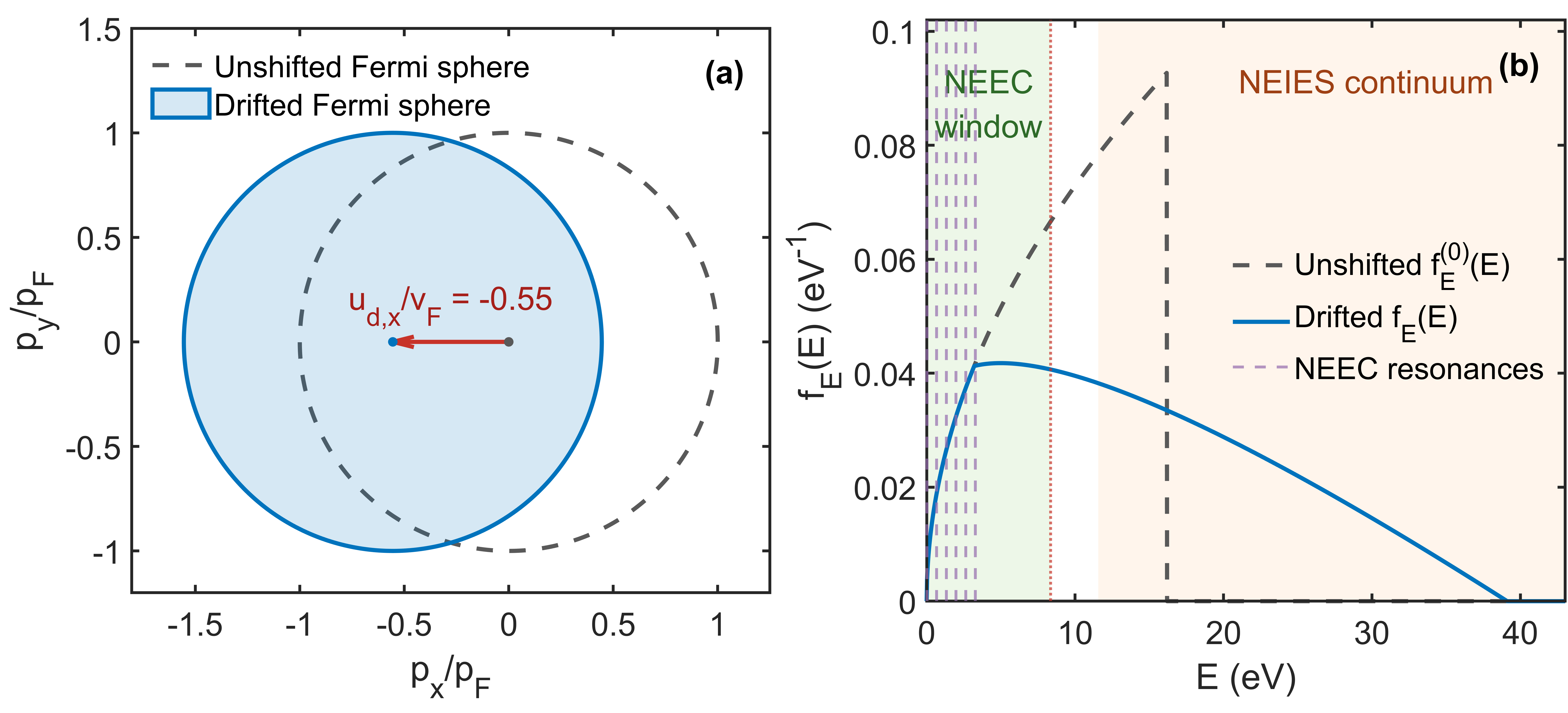}
		\caption{
			Local electron distribution at the inner wall and bunch center
			($r=a$, $\xi=0$), including density feedback.
			(a) Momentum-space sections of the unshifted and drifted Fermi
			spheres, with momenta normalized by the local Fermi momentum $p_F$.
			The return-current drift is opposite to the beam-propagation
			direction, with $u_{d,x}/v_F=-0.55$.
			(b) Corresponding unshifted and drifted energy distributions,
			$f_E^{(0)}(E)$ and $f_E(E)$. Shading indicates the NEEC window and
			the NEIES continuum. Vertical dashed lines mark representative NEEC
			resonances, and the red dotted line marks
			$E_{\mathrm{nuc}}=8.36~\mathrm{eV}$.
			The local Fermi and drift energies are
			$E_F=16.18~\mathrm{eV}$ and $E_d=4.98~\mathrm{eV}$, respectively.
		}
		\label{fig:fermi_spectrum}
	\end{figure}

	Figure~\ref{fig:fermi_spectrum} reveals the physical mechanism responsible for NEEC dominance. The beam-driven return current shifts the occupied Fermi sphere by the drift momentum $m_e\mathbf{u}_d$ without changing either its radius $p_F$ or the local electron density, as shown in Fig.~\ref{fig:fermi_spectrum}(a). Projection onto constant-energy shells produces the energy distribution $f_E(E)$ shown in Fig.~\ref{fig:fermi_spectrum}(b), with fully occupied states below $E_-=(\sqrt{E_F}-\sqrt{E_d})^2$ and partially occupied states extending to $E_+=(\sqrt{E_F}+\sqrt{E_d})^2$. The resulting distribution naturally supplies electrons at the discrete NEEC resonance energies \cite{PhysRevA.73.012715}. In the present case, NEIES requires incident electrons with $E>E_{\mathrm{nuc}}+E_-$, where $E_{\mathrm{nuc}}$ refers to the nuclear transition energy \cite{PhysRevLett.124.242501,PhysRevC.110.064621}. Electrons with $E_{\mathrm{nuc}}<E<E_{\mathrm{nuc}}+E_-$ cannot induce NEIES because the corresponding scattered-electron final states lie inside the occupied displaced Fermi sphere and are forbidden by Pauli blocking \cite{PhysRevResearch.1.033216}. Consequently, the same drifted Fermi distribution simultaneously provides resonant electrons for NEEC while suppressing NEIES through Pauli blocking, establishing the microscopic origin of the NEEC-dominant regime.
	
	The drifted electron spectrum enters the NEEC rate through the resonant electron population. For ion-like configuration $q$ and capture channel $\alpha$, the resonance energy is $E_{q\alpha}=E_{\mathrm{nuc}}-B_{q\alpha}$, where $B_{q\alpha}$ is the binding energy of the capture orbital. Using the integrated resonance strength $S_{q\alpha}=\int \sigma_{q\alpha}^{\mathrm{NEEC}}(E)\,dE$ \cite{PhysRevA.73.012715}, the local NEEC rate density is
	\begin{equation}
		R_{\mathrm{NEEC}}
		=
		n_i n_e
		\sum_q P_q
		\sum_{\alpha}
		S_{q\alpha}
		f_E(E_{q\alpha})
		v(E_{q\alpha}),
		\label{eq:neec_rate}
	\end{equation}
	where $n_i$ and $n_e$ are the local thorium and electron densities, and $v(E)$ is the electron velocity. The summation includes only energetically accessible capture channels with $E_{q\alpha}>0$. Equation~(\ref{eq:neec_rate}) shows that the NEEC rate is directly proportional to the population of resonant electrons through $f_E(E_{q\alpha})$.
	
	The NEIES rate is evaluated using the same electron distribution while accounting for Pauli blocking of the scattered-electron final states,
	\begin{equation}
		\begin{aligned}
			R_{\mathrm{NEIES}}
			={}&
			n_i n_e
			\sum_q P_q
			\int_{E_{\mathrm{nuc}}}^{\infty} dE\,
			f_E(E)
			\sigma_q^{\mathrm{NEIES}}(E)
			v(E)
			\\
			&\times
			\left[
			1-\bar f_d(E-E_{\mathrm{nuc}})
			\right],
		\end{aligned}
		\label{eq:neies_rate}
	\end{equation}
	where $\sigma_q^{\mathrm{NEIES}}(E)$ is the NEIES cross section \cite{PhysRevLett.124.242501,PhysRevC.110.064621}, and $\bar f_d(E-E_{\mathrm{nuc}})$ is the angularly averaged occupation probability of the final state. The factor $1-\bar f_d$ therefore represents the fraction of accessible final states. Unlike Eq.~(\ref{eq:neec_rate}), Eq.~(\ref{eq:neies_rate}) is reduced by Pauli blocking, which suppresses NEIES whenever the scattered-electron states remain occupied. The corresponding total excitation yields are obtained from $Y_j=\int dt\int dV\,R_j$ for $j=\mathrm{NEEC},\mathrm{NEIES}$.
	
	\textit{NEEC dominance and parameter dependence.---}
	For representative beam and target parameters, with $n_b=2.23\times10^{19}~\mathrm{cm^{-3}}$, $\sigma_x=3~\mu\mathrm{m}$, $\sigma_r=2~\mu\mathrm{m}$, $a=7~\mu\mathrm{m}$, and target length $L_{\mathrm{tube}}=1~\mathrm{mm}$, integrating the local excitation rates over the interaction volume and time yields $Y_{\mathrm{NEEC}}=2.40\times10^{5}$ and $Y_{\mathrm{NEIES}}=5.71\times10^{3}$. The corresponding NEEC fraction, $F_{\mathrm{NEEC}}\equiv Y_{\mathrm{NEEC}}/(Y_{\mathrm{NEEC}}+Y_{\mathrm{NEIES}})$, reaches $97.68\%$, establishing a NEEC-dominant excitation regime. Compared with the previously proposed laser-heated-cluster scheme, which predicts only $0.3-0.35$ NEEC excitations for a $10^{6}$-atom $^{229}\mathrm{Th}$ cluster \cite{PhysRevLett.130.112501}, the present approach increases the absolute excitation yield by several orders of magnitude through the combination of solid-density thorium and a self-organized return-current electron source.
	
	\begin{figure}[h]
		\centering
		\includegraphics[width=\columnwidth]
		{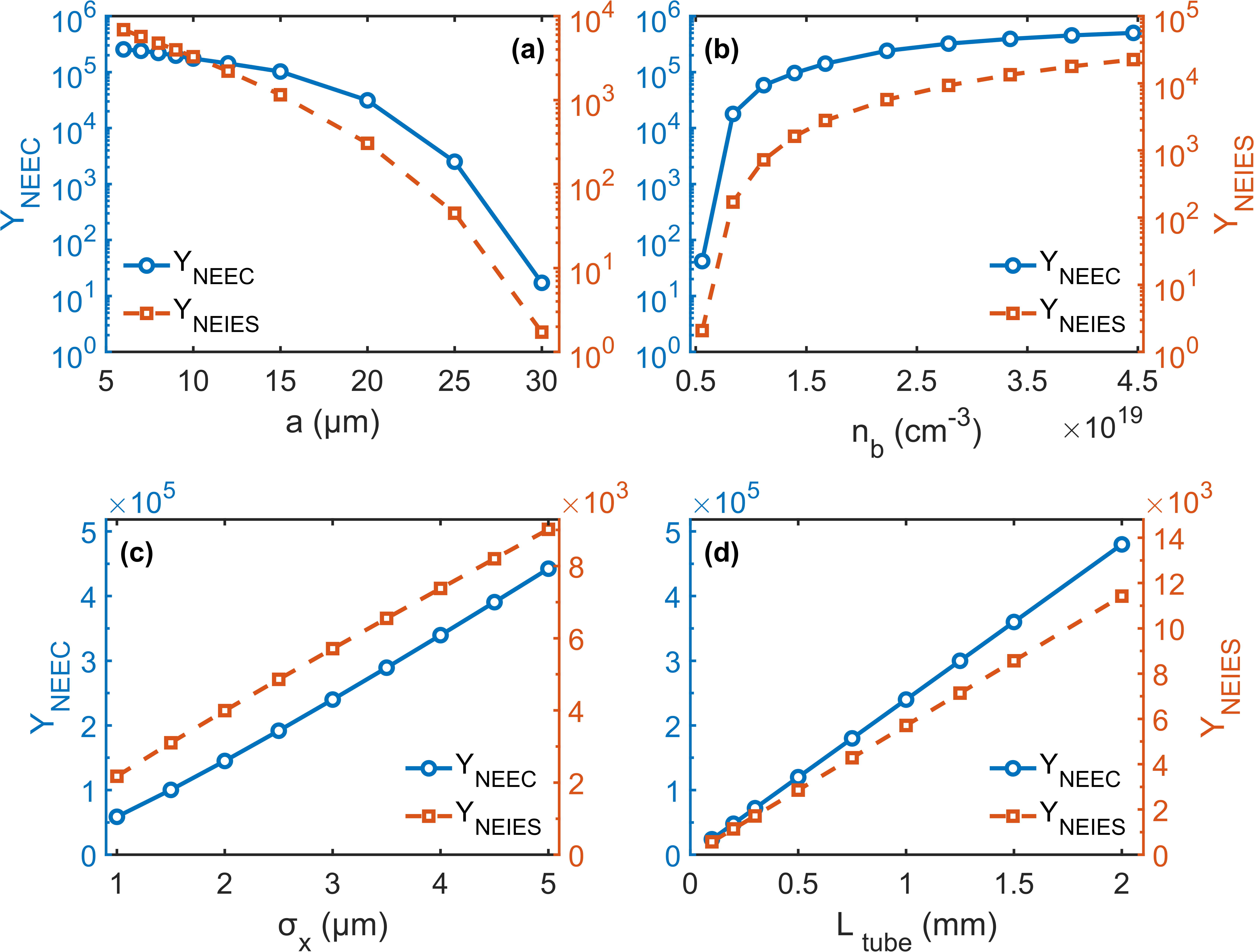}
		\caption{
			Dependence of the NEEC and NEIES yields on
			(a) the target inner radius $a$,
			(b) the peak beam density $n_b$,
			(c) the longitudinal rms bunch length $\sigma_x$, and
			(d) the target length $L_{\mathrm{tube}}$.
			Parameters not being varied are fixed at the values specified
			above.
		}
		\label{fig:param_scan}
	\end{figure}
	
	Figure~\ref{fig:param_scan} demonstrates that the NEEC-dominant regime is maintained over a broad range of beam and target parameters. The parameter dependence naturally separates into two categories. The first comprises the channel radius $a$ and beam density $n_b$, which directly determine the strength of the local return-current source. Increasing $a$ weakens the beam field at the inner wall and reduces both the drift energy and vacancy production, whereas increasing $n_b$ has the opposite effect, as shown in Figs.~\ref{fig:param_scan}(a) and (b). Consequently, the NEEC yield varies nonlinearly because the resonant electron population, vacancy generation, and ionic charge-state distribution evolve self-consistently. The second comprises the bunch length $\sigma_x$ and target length $L_{\mathrm{tube}}$, which primarily modify the interaction duration and volume when other parameters are fixed. Since the peak return-current density is nearly unchanged, increasing either parameter enhances the excitation yield mainly by extending the interaction time or increasing the number of participating nuclei. The resulting yield thus increases linearly with $\sigma_x$ and $L_{\mathrm{tube}}$, as indicated in Figs.~\ref{fig:param_scan}(c) and (d).
	
	Despite the large variations in the absolute excitation yield, the NEEC fraction remains remarkably stable. As shown in Fig.~\ref{fig:param_scan}, $F_{\mathrm{NEEC}}$ ranges from $90.9\%$ to $99.1\%$, while remaining within $97\%$-$99\%$ throughout the high-yield region. The lowest $F_{\mathrm{NEEC}}$ occurs only for the largest channel radius with $a=30~\mu\mathrm{m}$, where the weakened return current no longer generates capture vacancies efficiently and the NEEC yield has already decreased by more than four orders of magnitude.
	The weak variation of $F_{\mathrm{NEEC}}$ demonstrates that NEEC dominance is an intrinsic property of the beam-induced return-current electron source rather than a consequence of parameter optimization. Such robustness is particularly important for experiments because it relaxes the stringent requirements on beam and target optimization while preserving both a high NEEC yield and a dominant NEEC contribution.
	
	\textit{Discussion.---}
	The results above establish beam-induced surface return currents as an efficient and robust platform for NEEC-dominant nuclear excitation. We next discuss the validity of the underlying approximations, the robustness of the proposed mechanism, and also the experimental feasibility.
	
	(a) The theoretical model assumes that the return-current electrons remain well described by a drifted Fermi distribution. This approximation is justified because the collective surface response is established on the timescale of $\omega_p^{-1}\simeq $30-50$~\mathrm{as}$, which is much shorter than the beam duration $\tau_x\simeq10~\mathrm{fs}$. In the collisionless limit, the degenerate Fermi sea therefore follows the return-current drive through a rigid displacement in momentum space. Finite resistivity and electron collisions, however, can redistribute the electron energy through Ohmic heating and relax the drift motion \cite{PhysRevLett.42.890,PhysRevE.68.056404,PhysRevLett.97.235001}. To assess the influence of these effects, we replace the zero-temperature distribution by drifted Fermi-Dirac distributions with $k_BT_e=0$, $1$, and $2~\mathrm{eV}$, while reducing the drift energy to $30\%$-$100\%$ of its predicted value. Lower drift energies suppress impact ionization by reducing the high-energy electron population, whereas thermal broadening increases vacancy production but weakens Pauli blocking, thereby enhancing the competing NEIES channel. Nevertheless, recalculating the coupled vacancy dynamics and excitation rates still yields $F_{\mathrm{NEEC}}>95.86\%$ \cite{SupplementalMaterial}, demonstrating that the predicted NEEC-dominant regime is robust against moderate collisional relaxation and thermal broadening. A fully quantitative description will require a self-consistent treatment of collisional transport coupled to the beam-driven return current.
	
	(b) The predicted NEEC-dominant regime is also robust against uncertainties in the finite-density electronic structure. To assess the sensitivity to the capture-state binding energies, we apply a common shift $B_{q\alpha}\rightarrow B_{q\alpha}+\Delta B$ to all six capture levels and recalculate the coupled vacancy dynamics and excitation yields. Increasing $B_{q\alpha}$ ($\Delta B>0$) raises the impact-ionization thresholds while lowering the resonance energies $E_{q\alpha}=E_{\mathrm{nuc}}-B_{q\alpha}$, eventually closing channels with $E_{q\alpha}\leq0$. As $\Delta B$ varies from $-0.50$ to $1.00~\mathrm{eV}$, the NEEC yield $Y_{\mathrm{NEEC}}$ decreases from $2.67\times10^{5}$ to $1.12\times10^{5}$, while the NEEC fraction $F_{\mathrm{NEEC}}$ remains between $94.78\%$ and $97.99\%$. Even for $\Delta B=1.00~\mathrm{eV}$, where the two highest-binding-energy channels become energetically inaccessible, the remaining four channels continue to support a NEEC-dominant regime. These results demonstrate that the predicted NEEC dominance is not determined by the precise position of individual capture resonances \cite{SupplementalMaterial}.
	
	(c) Experimental realization of the proposed scheme appears feasible with current accelerator and target technologies. First of all, the representative beam parameters considered here lie within the operating range of facilities such as FACET-II \cite{PhysRevAccelBeams.22.101301}, and comparable beam parameters have also been demonstrated in state-of-the-art laser wakefield accelerators \cite{Couperus2017}. Secondly, the cylindrical geometry adopted here is chosen for analytical convenience. The underlying mechanism is not restricted to cylindrical channels. Beam-driven return currents are also well established in planar conducting targets \cite{PhysRevLett.109.255002}, suggesting that planar gaps, multichannel structures, and other metallic microstructured targets lined with $^{229}\mathrm{Th}$ should also support the proposed mechanism. Recent advances in the fabrication and application of microstructured targets further support the feasibility \cite{10.1063/1.5087409,Qin2026}. Thirdly, the resulting isomer population could be detected through delayed internal-conversion electrons. Recent time-gated measurements of delayed internal-conversion electrons from solid-state $^{229}\mathrm{ThO}_2$ \cite{Elwell2025} demonstrate the feasibility of separating the delayed nuclear signal from prompt electron backgrounds. In addition, the surface-return-current mechanism is not limited to $^{229}\mathrm{Th}$ and should be applicable to other low-energy nuclear transitions with accessible electron-capture resonances, such as the $76.7$-eV isomeric transition in $^{235}\mathrm{U}$ \cite{PhysRevC.110.L051601}.
	
	\textit{Conclusion.---}
	In conclusion, we have identified beam-induced surface return currents as a self-organized resonant electron source for NEEC. The return-current electrons simultaneously generate capture vacancies and provide resonant electrons, while Pauli blocking suppresses the competing NEIES channel, giving rise to a robust NEEC-dominant excitation regime. For experimentally accessible beam parameters, we predict $2.40\times10^{5}$ NEEC excitations per bunch with a NEEC fraction of $97.68\%$ in solid-density $^{229}\mathrm{Th}$. This regime persists over a broad range of beam and target parameters and remains robust against modeled variations in the electron distribution and finite-density electronic structure. By combining a high resonant-electron flux with efficient suppression of competing excitation channels, the proposed mechanism provides a practical route toward the long-sought experimental observation of dominant NEEC. More broadly, our work establishes beam-induced return currents as a new platform for resonant electron-driven nuclear excitation in solids.
	
	\begin{acknowledgments}
		This work was supported by the National Key R\&D Program of China (Grant Nos. 2022YFA1603300 and 2024YFA1613400), the National Natural Science Foundation of China (Grant Nos. 12375244, 12405288, and 12595364), the Natural Science Foundation of Hunan Province of China (Grant No. 2025JJ30002), the Hunan Provincial Innovation Foundation for Postgraduate (Grant No. CX20230008), the Innovation Program of Southwestern Institute of Physics (Grant No. 202501WDZCQN011), the Innovation Research Foundation of National University of Defense Technology (Grant Nos. XJQY2024046 and XJJC2024063), the Natural Science Foundation of Top Talent of SZTU (Grant No. GDRC202526), and the Shenzhen Science and Technology Program (Grant No. RCYX20221008092851073).
	\end{acknowledgments}
	
	\bibliographystyle{apsrev4-2}
	\bibliography{reference.bib}

	\end{document}


\title{Supplemental Material for ``Predominant Nuclear Excitation by Electron Capture Driven by Beam-Induced Return Currents''}

\author{Y. Y. Xu}
\thanks{These authors contributed equally to this work.}
\affiliation{College of Science, National University of Defense Technology,
	Changsha 410073, China}
\affiliation{Shenzhen Key Laboratory of Ultraintense Laser and Advanced Material Technology,
	Center for Intense Laser Application Technology, and College of Engineering Physics,
	Shenzhen Technology University, Shenzhen 518118, China}

\author{J. T. Qi}
\thanks{These authors contributed equally to this work.}
\affiliation{Shenzhen Key Laboratory of Ultraintense Laser and Advanced Material Technology,
	Center for Intense Laser Application Technology, and College of Engineering Physics,
	Shenzhen Technology University, Shenzhen 518118, China}

\author{H. Peng}
\affiliation{Shenzhen Key Laboratory of Ultraintense Laser and Advanced Material Technology,
	Center for Intense Laser Application Technology, and College of Engineering Physics,
	Shenzhen Technology University, Shenzhen 518118, China}

\author{K. Jiang}
\affiliation{Shenzhen Key Laboratory of Ultraintense Laser and Advanced Material Technology,
	Center for Intense Laser Application Technology, and College of Engineering Physics,
	Shenzhen Technology University, Shenzhen 518118, China}

\author{R. Li}
\affiliation{Shenzhen Key Laboratory of Ultraintense Laser and Advanced Material Technology,
	Center for Intense Laser Application Technology, and College of Engineering Physics,
	Shenzhen Technology University, Shenzhen 518118, China}

\author{Q. Xiao}
\affiliation{College of Science, National University of Defense Technology,
	Changsha 410073, China}

\author{T. P. Yu}
\email{tongpu@nudt.edu.cn}
\affiliation{College of Science, National University of Defense Technology,
	Changsha 410073, China}

\author{T. W. Huang}
\email{taiwu.huang@sztu.edu.cn}
\affiliation{Shenzhen Key Laboratory of Ultraintense Laser and Advanced Material Technology,
	Center for Intense Laser Application Technology, and College of Engineering Physics,
	Shenzhen Technology University, Shenzhen 518118, China}

\maketitle

\setcounter{equation}{0}
\renewcommand{\theequation}{S\arabic{equation}}

\setcounter{figure}{0}
\renewcommand{\thefigure}{S\arabic{figure}}

\setcounter{table}{0}
\renewcommand{\thetable}{S\Roman{table}}

\section{Return-current derivation and drifted Fermi spectrum}

Here we derive the electromagnetic response underlying the
return-current model used in the main text and project the displaced
Fermi occupation onto electron kinetic energy. The electron beam
propagates along the $+x$ direction, the inner channel radius is $a$,
and $\xi=t-x/c$ is the comoving coordinate. Throughout this section,
$Q$ denotes the magnitude of the bunch charge, while
$E_r^{\mathrm{vac}}$ and $B_\theta^{\mathrm{vac}}$ denote field
magnitudes.

For a Gaussian bunch with longitudinal rms duration
$\tau_x=\sigma_x/c$, the line-charge density is
\begin{equation}
\lambda(\xi)=
\frac{Q}{\sqrt{2\pi}\,c\tau_x}
\exp\!\left(-\frac{\xi^2}{2\tau_x^2}\right).
\end{equation}
For a transverse Gaussian profile with rms radius $\sigma_r$, the
ultrarelativistic near-field radial electric field is
\begin{equation}
E_r^{\mathrm{vac}}(r,\xi)=
\frac{\lambda(\xi)}{2\pi\varepsilon_0r}
\left[
1-\exp\!\left(-\frac{r^2}{2\sigma_r^2}\right)
\right],
\end{equation}
which applies for $r\ll\gamma\sigma_x$. When $a$ is larger than the
beam radius, the enclosed-charge factor approaches unity, giving
$E_r^{\mathrm{vac}}(a,\xi)\simeq\lambda(\xi)/(2\pi\varepsilon_0a)$
and
$B_\theta^{\mathrm{vac}}(a,\xi)
\simeq E_r^{\mathrm{vac}}(a,\xi)/c$ \cite{StupakovPenn2018}.

For a negatively charged beam propagating along $+x$, the electric
and magnetic fields point along $-\hat{\mathbf r}$ and
$-\hat{\boldsymbol\theta}$, respectively. The radial electric field
is screened primarily by surface charge, whereas the azimuthal
magnetic field penetrates the wall and drives an axial return current
\cite{PhysRevLett.107.135005}. In the collisionless cold-electron
limit, the axial current response satisfies
$\partial J_x/\partial t=\varepsilon_0\omega_p^2E_x$, where $n_e$
is the local delocalized-electron density and
$\omega_p=(e^2n_e/m_e\varepsilon_0)^{1/2}$ is the corresponding
electron plasma frequency. Using the Fourier
convention
$\widetilde F(\omega)=\int_{-\infty}^{\infty}
F(\xi)e^{-i\omega\xi}\,d\xi$, the incident magnetic field becomes
\begin{equation}
	\widetilde B_\theta^{\mathrm{vac}}(a,\omega)
	=
	\frac{Q}{2\pi\varepsilon_0c^2a}
	\exp\!\left(-\frac{\omega^2\tau_x^2}{2}\right).
\end{equation}

Because the fields depend on $x$ and $t$ only through $\xi$, each
Fourier component has the axial wavenumber $k_x=\omega/c$. In the
locally planar limit, for which the skin-layer thickness is much
smaller than the channel radius, the collisionless cold-plasma
dielectric function
$\varepsilon(\omega)=1-\omega_p^2/\omega^2$ gives the radial
attenuation constant
$\varkappa^2
=k_x^2-\varepsilon(\omega)\omega^2/c^2
=\omega_p^2/c^2$.
The corresponding collisionless skin depth is
$\delta=\varkappa^{-1}=c/\omega_p$, so that the magnetic field decays
inside the wall as
$\widetilde B_\theta(r,\omega)
=\widetilde B_\theta(a,\omega)
\exp[-(r-a)/\delta]$. The accompanying radial electric-field
amplitude satisfies
$
\frac{|\widetilde E_r|}
{c|\widetilde B_\theta|}
=
\frac{\omega^2}{\omega_p^2-\omega^2}
\simeq
\left(\frac{\omega}{\omega_p}\right)^2 .
$
Thus, for the characteristic bunch frequency
$\omega\sim\tau_x^{-1}\ll\omega_p$, the radial electric field inside
the wall is strongly screened.

Ampère's law relates the radial gradient of the magnetic field to the
axial current density. Neglecting displacement-current corrections of
relative order $(\omega/\omega_p)^2$ and transforming back to $\xi$
gives
\begin{equation}
	J_x(r,\xi)=
	\frac{\omega_pQ}{(2\pi)^{3/2}c\tau_xa}
	\exp\!\left(-\frac{\xi^2}{2\tau_x^2}\right)
	\exp\!\left[-\frac{r-a}{\delta}\right].
	\label{eq:S_return_current}
\end{equation}
Here $J_x>0$ denotes the current along $+x$, while the corresponding
electron drift is along $-x$. For a nonuniform electron density, the local-density approximation
gives
\begin{equation}
J_x(r,\xi)
\simeq
\frac{B_\theta^{\mathrm{vac}}(a,\xi)}
{\mu_0\delta(r,\xi)}
\exp\!\left[
-\int_a^r\frac{dr'}{\delta(r',\xi)}
\right],
\end{equation}
where $\delta(r,\xi)=c/\omega_p(r,\xi)$. This form also preserves the
beam-imposed sheet current.

The longitudinal electric field associated with the buildup and decay
of the return current follows from
$E_x=(\varepsilon_0\omega_p^2)^{-1}\partial J_x/\partial\xi$.
For the fixed-density cold response used in the field-ionization
estimate,
\begin{equation}
	E_x(r,\xi)=
	-\frac{Q\xi}
	{(2\pi)^{3/2}\varepsilon_0c\omega_p\tau_x^3a}
	\exp\!\left(-\frac{\xi^2}{2\tau_x^2}\right)
	\exp\!\left[-\frac{r-a}{\delta}\right].
	\label{eq:S_longitudinal_field}
\end{equation}
The field is antisymmetric about the bunch center and is used only in
the field-assisted-ionization check below.

The return current produces an electron drift
$\mathbf u_d=-u_d\hat{\mathbf x}$, with
$u_d=J_x/(en_e)$ and $E_d=m_eu_d^2/2$. At zero temperature,
$p_F=\hbar(3\pi^2n_e)^{1/3}$ and $E_F=p_F^2/(2m_e)$ \cite{AshcroftMermin1976}. The displaced
momentum-space occupation is
\begin{equation}
	f_d(\mathbf p)=
	\Theta\!\left[
	p_F-\left|\mathbf p-m_e\mathbf u_d\right|
	\right].
	\label{eq:S_drifted_occupation}
\end{equation}
To project this occupation onto kinetic energy
$E=p^2/(2m_e)$, let
$\mu=\hat{\mathbf p}\cdot\hat{\mathbf u}_d$. The occupation condition
is $\mu\geq\mu_c$, where
$\mu_c=(E+E_d-E_F)/(2\sqrt{EE_d})$. The occupied fraction of a
constant-energy shell is
\begin{equation}
	\mathcal{A}(E;E_F,E_d)=
	\begin{cases}
		1, & \mu_c\leq-1,\\[2pt]
		(1-\mu_c)/2, & -1<\mu_c<1,\\[2pt]
		0, & \mu_c\geq1.
	\end{cases}
	\label{eq:S_angular_fraction}
\end{equation}
Multiplying the free-electron density of states by this angular
fraction and normalizing by $n_e$ gives
\begin{equation}
	f_E(E;r,\xi)=
	\frac{3\sqrt{E}}{2E_F^{3/2}}\,
	\mathcal{A}(E;E_F,E_d).
	\label{eq:S_energy_distribution}
\end{equation}
It satisfies $\int_0^\infty f_E(E;r,\xi)\,dE=1$. In the limit
$E_d\rightarrow0$, it reduces to
$f_E^{(0)}(E)=3\sqrt{E}\,
\Theta(E_F-E)/(2E_F^{3/2})$.

The boundaries of the projected spectrum are
$E_\pm=(\sqrt{E_F}\pm\sqrt{E_d})^2$. For $E_d<E_F$, the energy
shells are fully occupied below $E_-$, partially occupied between
$E_-$ and $E_+$, and empty above $E_+$. The angle-averaged occupation
of a scattered-electron final state is consequently
$
\bar f_d(E_f;r,\xi)
=
\mathcal{A}\!\left[
E_f;E_F(r,\xi),E_d(r,\xi)
\right],
$
which is used in the Pauli-blocking factor of the NEIES rate.

\section{Metallic thorium and vacancy production}

At the solid density of thorium,
$\rho_0=11.7~\mathrm{g\,cm^{-3}}$, we calculate the electronic
structure using the finite-density average-atom code atoMEC
\cite{PhysRevResearch.4.023055,callow2022}. The average-atom model
replaces the many-ion system by a representative atom embedded in a
charge-neutral spherical cell at the prescribed density and
temperature. Within finite-temperature density-functional theory,
atoMEC self-consistently determines the bound and continuum-like
electronic states, including the effects of plasma screening on their
energies and occupations. The calculation is performed at
$k_BT_e=0.1~\mathrm{eV}$, representing the strongly degenerate
metallic state used in the excitation model. The resulting
eigenvalues and integrated occupations are summarized in
Table~\ref{tab:S_atomec_bands}.

\begin{table}[h]
	\centering
	\caption{
		Representative finite-density atoMEC eigenvalues and integrated
		occupations for solid-density thorium at
		$k_BT_e=0.1~\mathrm{eV}$. Energies are measured relative to the
		continuum edge.
	}
	\label{tab:S_atomec_bands}
	\begin{tabular}{c c c c}
		\hline\hline
		State & Character & Eigenvalue range (eV) & Occupation $N_e$ \\
		\hline
		$5p$ & localized bound & $-157.21$ & $6.000$ \\
		$5d$ & localized bound & $-77.14$ & $10.000$ \\
		$6s$ & localized bound & $-19.37$--$-18.17$ & $1.986$ \\
		$6p$ & localized bound & $-8.34$--$-5.07$ & $5.959$ \\
		$6d$ & continuum-like & $4.88$--$8.38$ & $4.052$ \\
		$5f$ & continuum-like & $8.44$--$12.79$ &
		$1.99\times10^{-3}$ \\
		\hline
		All bound states & localized bound & -- & $85.946$ \\
		All continuum-like states & continuum-like & -- & $4.054$ \\
		\hline\hline
	\end{tabular}
\end{table}

The localized population is close to the 86-electron
$[\mathrm{Rn}]$ closed-shell core, while the integrated
continuum-like occupation corresponds to approximately four
delocalized electrons per thorium atom. We therefore represent the
initial metallic state by an effective ion-like configuration
$q=4$, together with a delocalized-electron density $n_e=4n_i$.
Configurations with $q>4$ describe additional localized vacancies
produced by the return-current electrons.

The localized $6p$ states provide the finite-density binding energies
used in both the impact-ionization model and the NEEC resonance
condition. For the sequential transitions $q\rightarrow q+1$ with
$q=4$--$9$, we use
\begin{equation}
	B_q=
	\{5.07,\,5.72,\,6.38,\,7.03,\,7.68,\,8.34\}
	~\mathrm{eV},
	\label{eq:S_6p_binding_energies}
\end{equation}
respectively. These values define the six successive
$6p$-vacancy-production thresholds. The corresponding
fine-structure-resolved capture binding energies $B_{q\alpha}$ enter
the NEEC resonance condition below.

The single-particle $6s$ eigenvalues in
Table~\ref{tab:S_atomec_bands} characterize the initial metallic
configuration and do not represent the sequential ionization
threshold after all six $6p$ electrons have been removed. We therefore estimate the onset of deeper-shell ionization using the
high-density (ion-sphere) limit of the Stewart--Pyatt
continuum-lowering model \cite{STEWART19661203}. For the first
$6s$ ionization step, $q=10\rightarrow11$, this gives
$I_{10}^{\mathrm{eff}}=45.47~\mathrm{eV}$.

At the inner wall and bunch center, $r=a$ and $\xi=0$, where the
return-current drive is close to its maximum, the representative
calculation gives $E_F=16.18~\mathrm{eV}$ and
$E_d=4.98~\mathrm{eV}$. The upper spectral edge
$E_+=(\sqrt{E_F}+\sqrt{E_d})^2$ therefore gives
$E_+-E_F=22.93~\mathrm{eV}$, well below the estimated first
$6s$ threshold. The ionization cascade consequently remains confined
to $q\leq10$. In the numerical calculation, the state space is
nevertheless extended to $q=20$ using the Stewart--Pyatt estimates
for the deeper-shell thresholds as a cutoff check; these higher
configurations remain negligibly populated and do not affect the
excitation yields.

All quantities below are local, and their dependence on $(r,\xi)$ is
left implicit. Let
$\mathbf{P}=(P_4,P_5,\ldots,P_{20})^{\mathsf T}$ denote the
configuration fractions, with $P_4=1$ before the arrival of the
bunch. Sequential vacancy production obeys
\begin{equation}
	\frac{\partial\mathbf{P}}{\partial\xi}
	=
	\mathbf{A}(\xi)\mathbf{P},
	\label{eq:S_population_dynamics}
\end{equation}
where $\mathbf{A}$ transfers population from configuration $q$ to
$q+1$ at the total sequential-ionization rate $\Gamma_q$ defined
below.

The impact-ionization contribution is evaluated using modified
binary-encounter Bethe (MBEB) cross sections
\cite{GUERRA20121}. Since the equilibrium Fermi sea does not
constitute a return-current-driven collision source, we retain only
the positive nonequilibrium increment
$
\Delta f_+(E)
=
\max\!\left[
f_E(E)-f_E^{(0)}(E),0
\right].
$
Defining the kinetic energy above the occupied Fermi sea as
$\mathcal{E}=E-E_F$, the rate for $q\rightarrow q+1$ is
\begin{equation}
	\Gamma_q^{\mathrm{MBEB}}
	=
	n_e
	\int_{B_q}^{\infty}d\mathcal{E}\,
	\Delta f_+(E_F+\mathcal{E})
	v(\mathcal{E})
	\sigma_q^{\mathrm{MBEB}}(\mathcal{E}).
	\label{eq:S_MBEB_rate}
\end{equation}
The lower limit enforces $\mathcal{E}\geq B_q$.
The increment $\Delta f_+$ is not renormalized, so its integral gives
the drift-induced electron fraction in the high-energy part of the
distribution.

Each ionization step adds one electron to the delocalized population,
giving
$n_e=n_i\sum_{q=4}^{20}qP_q$. The resulting density feedback updates
$E_F$ and $\omega_p$. Preserving the beam-imposed sheet current,
$
\int_a^\infty J_x\,dr=\frac{I_b}{2\pi a},
$
the corresponding change in the skin depth redistributes $J_x$,
which in turn updates $E_d$ and the electron distribution.

Field-assisted ionization driven by the longitudinal field $E_x$ is
included only as a consistency check. The total sequential-ionization
rate is
$
\Gamma_q
=
\Gamma_q^{\mathrm{MBEB}}
+
\Gamma_q^{\mathrm{FI}}.
$
The contribution $\Gamma_q^{\mathrm{FI}}$ is estimated using an
effective ADK tunneling model, supplemented by the Bauer--Mulser and
barrier-suppression limits at higher fields
\cite{Ammosov1986ADK,PhysRevA.59.569,PhysRevA.98.043407},
with the same effective ionization thresholds as in the MBEB
sequence.

For the representative parameters of the main text,
$|E_x|_{\max}=7.11\times10^8~\mathrm{V/m}$, while a conservative
upper bound on the cumulative field-ionization probability over the
complete bunch passage is only $4.00\times10^{-28}$.
Field-assisted ionization is therefore negligible, and the
configuration dynamics are governed by return-current-driven MBEB
impact ionization.

\section{NEEC and NEIES rates and yields}

The nuclear-excitation rates are evaluated using the local electron
distribution and ion-like populations defined above. For an
ion-like configuration $q$ and capture channel $\alpha$, the NEEC
resonance energy is
$E_{q\alpha}=E_{\mathrm{nuc}}-B_{q\alpha}$,
where $E_{\mathrm{nuc}}=8.35574~\mathrm{eV}$ and $B_{q\alpha}$ is
the finite-density binding energy of the capture orbital. Channels
with $E_{q\alpha}\leq0$ are energetically closed.

The NEEC-active configurations are $q=5,\ldots,10$, corresponding to
one to six dynamically produced $6p$ vacancies. Configurations
$q=5,\ldots,8$ contain vacancies in the $6p_{3/2}$ subshell, while
$q=9$ and $10$ also contain $6p_{1/2}$ vacancies. The channel set
includes the allowed fine-structure-resolved contributions from the
M1 and E2 components of the $^{229}\mathrm{Th}$ nuclear transition.

For a given channel, the NEEC cross section is written as
\cite{PhysRevA.73.012715}
\begin{equation}
	\begin{aligned}
		\sigma_{q\alpha}^{\mathrm{NEEC}}(E)
		={}&
		\frac{4\pi^2}{c^2}
		\frac{\epsilon_i}{p_i^3}
		\sum_{\tau\lambda}
		\frac{
			B(\tau\lambda;I_g\rightarrow I_m)
			\kappa^{2\lambda+2}
		}{
			[(2\lambda+1)!!]^2
		}
		\\
		&\times
		\sum_{\eta_i}
		(2j_i+1)
		\left|C_{j_i j_f}^{(\lambda)}\right|^2
		\left|M_{q\alpha}^{\tau\lambda}\right|^2
		\frac{\Gamma_{q\alpha}^{\mathrm{NEEC}}}
		{
			(E-E_{q\alpha})^2+
			(\Gamma_{q\alpha}^{\mathrm{NEEC}})^2/4
		}.
	\end{aligned}
	\label{eq:S_NEEC_cross_section}
\end{equation}
Here $E$ is the incident electron kinetic energy,
$\epsilon_i=E+m_ec^2$,
$p_i=c^{-1}(\epsilon_i^2-m_e^2c^4)^{1/2}$,
$\tau\lambda=\mathrm{M1},\mathrm{E2}$, and
$\kappa=E_{\mathrm{nuc}}/c$.
The quantity $B(\tau\lambda;I_g\rightarrow I_m)$ is the reduced
nuclear transition probability,
$C_{j_i j_f}^{(\lambda)}$ is the angular coefficient, and
$M_{q\alpha}^{\tau\lambda}$ is the radial matrix element of the
free-bound electronic transition.

Because the resonance width is much narrower than the variation
scale of the electron spectrum, each channel is characterized by its
integrated resonance strength
$S_{q\alpha}=\int dE\,\sigma_{q\alpha}^{\mathrm{NEEC}}(E)$.
The local NEEC rate density is then
\begin{equation}
	R_{\mathrm{NEEC}}
	=
	n_i n_e
	\sum_{q=5}^{10}P_q
	\sum_{\alpha\in\mathcal{C}_q}
	S_{q\alpha}
	f_E(E_{q\alpha})
	v(E_{q\alpha}),
	\label{eq:S_NEEC_rate}
\end{equation}
where $\mathcal{C}_q$ is the set of energetically open channels and
$v(E)$ is the electron velocity.

\begin{figure}[h]
	\centering
	\includegraphics[width=0.73\textwidth]
	{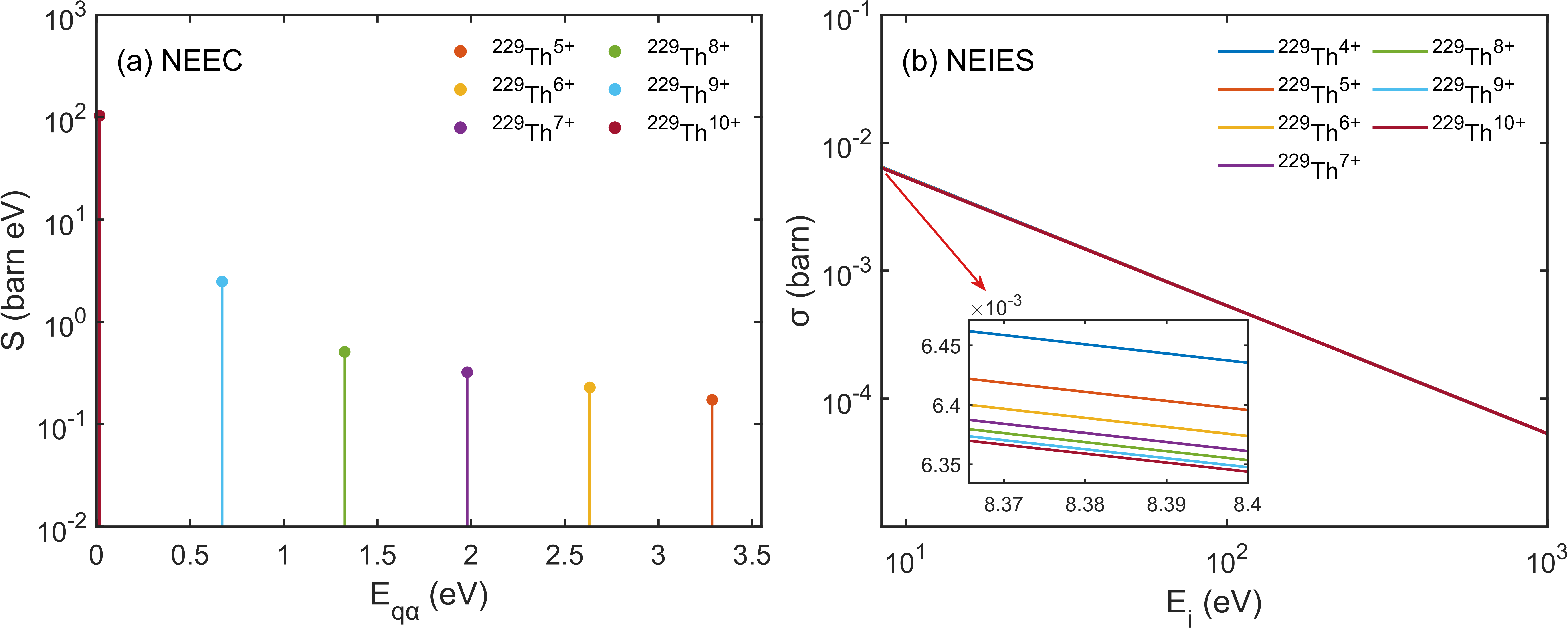}
	\caption{
		Charge-state-resolved nuclear-excitation quantities used in
		the rate calculations.
		(a) Integrated NEEC resonance strengths $S_{q\alpha}$ at the
		corresponding resonance energies for $q=5,\ldots,10$.
		(b) NEIES cross sections $\sigma_q^{\mathrm{NEIES}}$ as
		functions of the incident-electron kinetic energy $E_e$ for
		$q=4,\ldots,10$. The inset resolves their near-threshold
		charge-state dependence.
	}
	\label{fig:S_cross_sections}
\end{figure}

For NEIES, an incident electron with kinetic energy
$E_i\geq E_{\mathrm{nuc}}$ remains in the continuum with final
kinetic energy $E_f=E_i-E_{\mathrm{nuc}}$. The corresponding cross
section is
\cite{PhysRevLett.124.242501,PhysRevC.106.044604,
	PhysRevC.110.064621}
\begin{equation}
	\begin{aligned}
		\sigma_q^{\mathrm{NEIES}}(E_i)
		={}&
		\frac{8\pi^2}{c^4}
		\frac{\epsilon_f}{p_f}
		\frac{\epsilon_i}{p_i^3}
		\sum_{\tau\lambda}
		\frac{
			B(\tau\lambda;I_g\rightarrow I_m)
			\kappa^{2\lambda+2}
		}{
			[(2\lambda+1)!!]^2
		}
		\\
		&\times
		\sum_{\eta_i,\eta_f}
		(2j_i+1)
		\left|C_{j_i j_f}^{(\lambda)}\right|^2
		\left|
		M_q^{\tau\lambda}(E_i,E_f)
		\right|^2 .
	\end{aligned}
	\label{eq:S_NEIES_cross_section}
\end{equation}
Here $\epsilon_{i,f}=E_{i,f}+m_ec^2$ and
$p_{i,f}=c^{-1}(\epsilon_{i,f}^2-m_e^2c^4)^{1/2}$.
The quantity $M_q^{\tau\lambda}(E_i,E_f)$ is the radial matrix
element of the free-free electronic transition in the ion-core
potential. The resulting integrated NEEC resonance strengths and NEIES cross
sections are summarized in Fig.~\ref{fig:S_cross_sections}.

The local NEIES rate density, including final-state Pauli blocking,
is
\begin{equation}
	\begin{aligned}
		R_{\mathrm{NEIES}}
		={}&
		n_i n_e
		\sum_{q\in\mathcal{Q}}P_q
		\int_{E_{\mathrm{nuc}}}^{\infty}dE_i\,
		f_E(E_i)
		\sigma_q^{\mathrm{NEIES}}(E_i)
		v(E_i)
		\\
		&\times
		\left[
		1-\bar f_d(E_i-E_{\mathrm{nuc}})
		\right].
	\end{aligned}
	\label{eq:S_NEIES_rate}
\end{equation}
The factor $1-\bar f_d(E_i-E_{\mathrm{nuc}})$ is the
angle-averaged fraction of available final states for the scattered
electron \cite{PhysRevResearch.1.033216}. The set $\mathcal{Q}$ contains the populated ion-like
configurations included in the NEIES calculation.

\begin{figure}[h]
	\centering
	\includegraphics[width=0.73\textwidth]
	{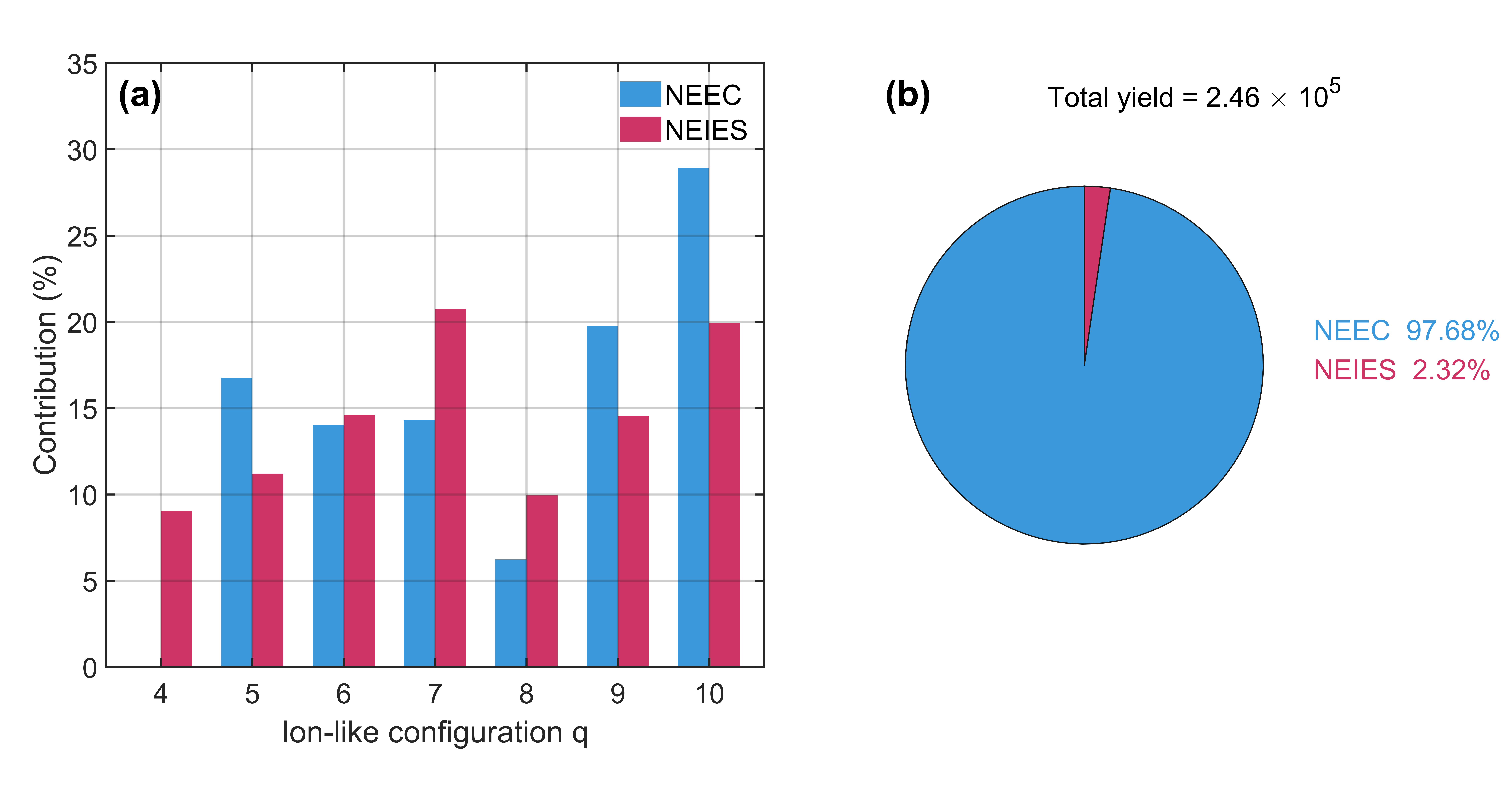}
	\caption{
		NEEC and NEIES yield decomposition for the representative
		parameters of the main text.
		(a) Configuration-resolved fractional contributions to
		$Y_{\mathrm{NEEC}}$ and $Y_{\mathrm{NEIES}}$, normalized
		separately for the two mechanisms.
		(b) NEEC and NEIES fractions of their combined
		electron-induced excitation yield.
	}
	\label{fig:S_yield_decomposition}
\end{figure}

For a target of length $L_{\mathrm{tube}}$ with a uniform local
response along its axis, the yield in channel $X$ is
\begin{equation}
	Y_X
	=
	L_{\mathrm{tube}}
	\int d\xi
	\int 2\pi r\,dr\,
	R_X(r,\xi),
	\qquad
	X=\mathrm{NEEC},\mathrm{NEIES}.
	\label{eq:S_yield_integration}
\end{equation}
At a fixed snapshot, $\xi=t-x/c$, so $d\xi=-dx/c$. On the numerical
$(x,r)$ grid, the yield is therefore evaluated as
$
Y_X
\simeq
\frac{L_{\mathrm{tube}}}{c}
\sum_{i,j}
R_X(r_j,\xi_i)\,
2\pi r_j\,\Delta r\,|\Delta x|$.

The factor $L_{\mathrm{tube}}/c$ accounts for the accumulation of the
same local comoving response along the target length; the exposure
time at any fixed axial position remains determined by the bunch
envelope.

For the representative parameter set, the calculation gives
$Y_{\mathrm{NEEC}}=2.40\times10^5$ and
$Y_{\mathrm{NEIES}}=5.71\times10^3$.
Figure~\ref{fig:S_yield_decomposition}(a) shows that the initial
$q=4$ configuration contributes only to NEIES because it contains no
$6p$ capture vacancy. NEEC begins at $q=5$ and receives contributions
from all six vacancy-bearing configurations, with the largest
fractions arising from $q=9$ and $10$. This distribution reflects the
combined effects of the configuration populations and the number and
strengths of the available $6p$ capture channels. In contrast, NEIES
does not require a capture vacancy and is distributed more broadly
over the populated configurations, although its continuum final-state
phase space is restricted by Pauli blocking. The combined yield is
$Y_{\mathrm{tot}}=2.46\times10^5$, of which
$F_{\mathrm{NEEC}}=97.68\%$, as shown in
Fig.~\ref{fig:S_yield_decomposition}(b).

\section{Sensitivity to electron heating and drift relaxation}

The collective electronic response is established on the plasma
timescale $\omega_p^{-1}\simeq0.03$--$0.05~\mathrm{fs}$, which is
much shorter than the bunch duration $\tau_x\simeq10~\mathrm{fs}$.
The local momentum distribution can therefore follow the
return-current drive on the bunch-envelope timescale. In the
collisionless limit, a locally uniform force translates the
degenerate Fermi sea in momentum space, providing the basis for the
displaced-Fermi description used above. Finite resistivity and
collisional transport can nevertheless produce Ohmic heating,
redistribute the electron energy, and relax the directed drift
\cite{PhysRevLett.42.890,PhysRevE.68.056404,
	PhysRevLett.97.235001}. Because these processes are not evolved
dynamically in the reduced return-current model, we assess their
influence by varying the electron temperature and scaling the local
drift energy as
\begin{equation}
	E_d^{(\eta)}
	=
	\eta_{E_d}E_d,
	\qquad
	0.3\leq\eta_{E_d}\leq1,
	\label{eq:S_drift_reduction}
\end{equation}
where $E_d$ is the value predicted by the return-current model. The
corresponding drift velocity is
$\mathbf{u}_d^{(\eta)}
=\sqrt{\eta_{E_d}}\,\mathbf{u}_d$.

At finite temperature, the displaced Fermi occupation becomes
\begin{equation}
	f_d(\mathbf{p})
	=
	\left\{
	\exp\!\left[
	\frac{
		|\mathbf{p}-m_e\mathbf{u}_d^{(\eta)}|^2/(2m_e)
		-\mu_e
	}{
		k_BT_e
	}
	\right]
	+1
	\right\}^{-1},
	\label{eq:S_drifted_FD}
\end{equation}
where $\mu_e$ is fixed by the local electron density. Projecting this
occupation onto a kinetic-energy shell gives
\begin{equation}
	f_E(E)
	=
	\frac{D(E)}{2n_e}
	\int_{-1}^{1}
	\frac{d\chi}{
		\exp\!\left[
		\bigl(
		E+E_d^{(\eta)}
		-2\sqrt{E E_d^{(\eta)}}\,\chi
		-\mu_e
		\bigr)/(k_BT_e)
		\right]
		+1
	},
	\label{eq:S_finite_T_spectrum}
\end{equation}
where
$D(E)=(2\pi^2)^{-1}(2m_e/\hbar^2)^{3/2}\sqrt{E}$ and
$\chi=\hat{\mathbf p}\cdot\hat{\mathbf u}_d^{(\eta)}$.
The chemical potential is fixed by particle-number conservation,
$\int_0^\infty D(E)\bar f_d(E)\,dE=n_e$. The angularly averaged occupation $\bar f_d(E)$ used for NEIES is
given by the same angular integral without the prefactor
$D(E)/n_e$. In the limit $T_e\rightarrow0$, these expressions reduce
to the zero-temperature distribution derived above.

We consider $k_BT_e=0$, $1$, and $2~\mathrm{eV}$ together with
$\eta_{E_d}=0.3$, $0.5$, $0.7$, and $1$. For $T_e=0$, the
zero-temperature expressions are used directly. For every parameter pair, the vacancy dynamics, density feedback,
and both excitation rates are recalculated, including the
corresponding final-state Pauli-blocking factor for NEIES.

\begin{figure}[h]
	\centering
	\includegraphics[width=0.73\textwidth]
	{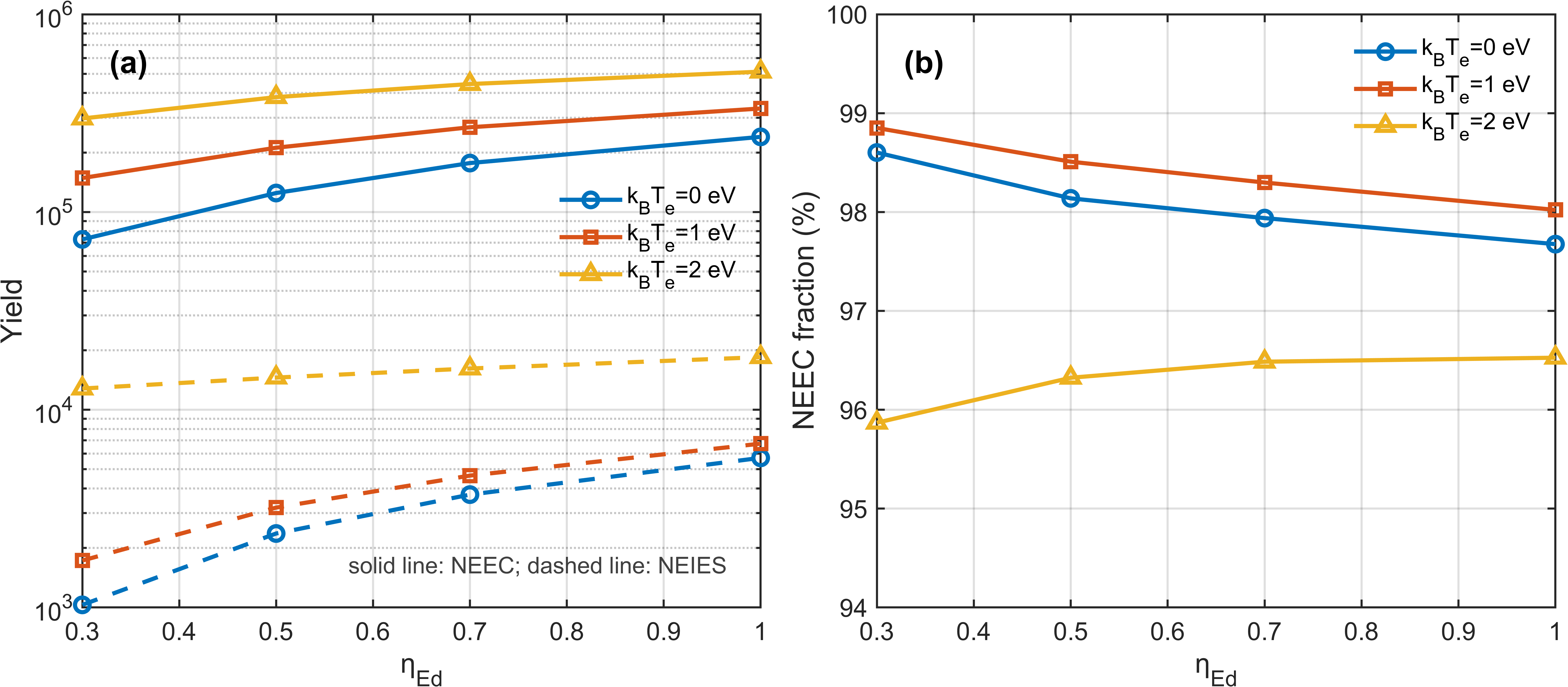}
	\caption{
		Sensitivity of the excitation yields to electron heating and
		drift relaxation.
		(a) NEEC (solid lines) and NEIES (dashed lines) yields as functions of the drift-energy factor $\eta_{E_d}$ for
		$k_BT_e=0$, $1$, and $2~\mathrm{eV}$.
		(b) Corresponding NEEC fraction
		$F_{\mathrm{NEEC}}
		=Y_{\mathrm{NEEC}}/
		(Y_{\mathrm{NEEC}}+Y_{\mathrm{NEIES}})$.
		All beam and target parameters are fixed at the representative
		values used in the main text.
	}
	\label{fig:S_thermal_drift_sensitivity}
\end{figure}

As shown in Fig.~\ref{fig:S_thermal_drift_sensitivity}(a), both
yields increase with $\eta_{E_d}$. A larger retained drift enhances
the high-energy population responsible for vacancy production and
NEIES, while also increasing the electron population sampled at the
NEEC resonance energies. Increasing the temperature also raises the absolute
yields by broadening the high-energy part of the distribution.
Its relative effect on the two mechanisms, however, is different.

For $k_BT_e=0$ and $1~\mathrm{eV}$, the fractional increase of
$Y_{\mathrm{NEIES}}$ with $\eta_{E_d}$ is slightly larger than that
of $Y_{\mathrm{NEEC}}$, producing the modest decrease of
$F_{\mathrm{NEEC}}$ in
Fig.~\ref{fig:S_thermal_drift_sensitivity}(b). At
$k_BT_e=2~\mathrm{eV}$, thermal broadening already weakens
final-state Pauli blocking at low drift energy; increasing
$\eta_{E_d}$ then produces a stronger relative increase in NEEC, so
the NEEC fraction rises slightly and approaches a plateau. Across the
full scan, $F_{\mathrm{NEEC}}$ remains between $95.87\%$ and
$98.85\%$, with the minimum at $k_BT_e=2~\mathrm{eV}$ and
$\eta_{E_d}=0.3$. The predominance of NEEC therefore does not depend
on an unheated or fully retained drift distribution. A quantitative time
history would require a collisional-transport model coupled to the
driven return current.

\section{Sensitivity to the finite-density $6p$ binding energies}

The finite-density $6p$ binding energies enter the calculation in two
ways: they determine the NEEC resonance energies through
$E_{q\alpha}=E_{\mathrm{nuc}}-B_{q\alpha}$ and set the MBEB
thresholds for the sequential production of $6p$ vacancies. To test
the sensitivity to a systematic offset in these energies, we apply
the common shift
\begin{equation}
	B_q\rightarrow B_q+\Delta B,
	\qquad
	q=4,\ldots,9,
	\label{eq:S_B_shift}
\end{equation}
and apply the same shift to the corresponding capture binding
energies $B_{q\alpha}$. For each value of $\Delta B$, the vacancy
populations, electron-density feedback, NEEC resonance energies, and
NEEC and NEIES yields are recalculated. Channels satisfying
$E_{\mathrm{nuc}}-B_{q\alpha}\leq0$ are excluded by the same energy
condition as in the baseline calculation.

\begin{figure}[h]
	\centering
	\includegraphics[width=0.7\textwidth]
	{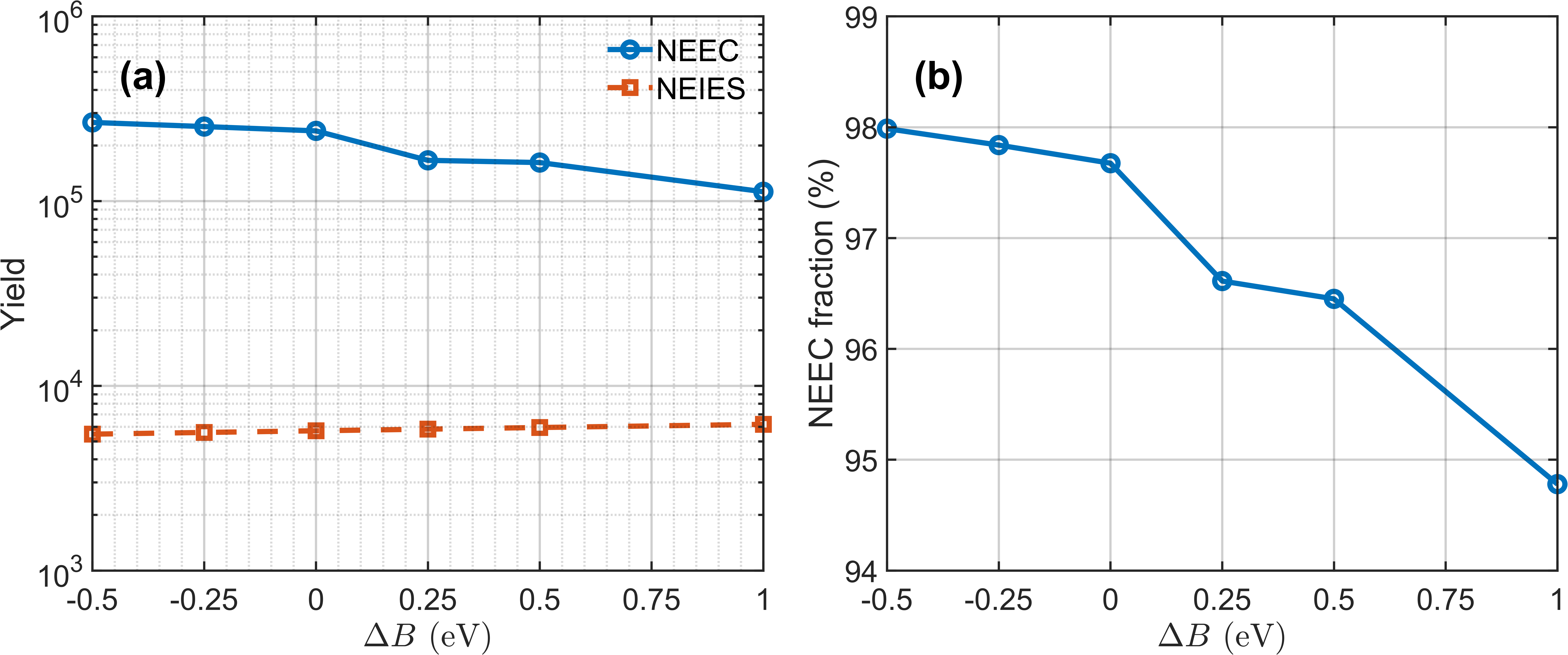}
	\caption{
		Sensitivity of the NEEC and NEIES yields to a common shift
		$\Delta B$ of the finite-density $6p$ binding energies.
		(a) NEEC and NEIES yields.
		(b) Corresponding NEEC fraction.
	}
	\label{fig:S_B_sensitivity}
\end{figure}

The results are summarized in
Fig.~\ref{fig:S_B_sensitivity} and
Table~\ref{tab:S_B_sensitivity}. The binding-energy shift affects
NEEC through two coupled mechanisms.
A positive $\Delta B$ raises the MBEB thresholds and therefore
suppresses the production of $6p$ vacancies. At the same time, it
lowers the resonance energies
$E_{q\alpha}=E_{\mathrm{nuc}}-B_{q\alpha}$ and closes a capture
channel when $B_{q\alpha}\geq E_{\mathrm{nuc}}$.

For $\Delta B<0$, all six vacancy-bearing configurations retain
energetically open NEEC channels, while the reduced ionization
thresholds increase their vacancy populations and hence the NEEC
yield. At $\Delta B=+0.25~\mathrm{eV}$, the $q=10$ NEEC channel,
which has the largest capture binding energy, becomes energetically
closed. It remains closed at $\Delta B=+0.50~\mathrm{eV}$, so five
configurations contribute to NEEC in both cases. At
$\Delta B=+1.00~\mathrm{eV}$, the $q=9$ NEEC channel also closes,
leaving four configurations that contribute to NEEC.

As shown in Fig.~\ref{fig:S_B_sensitivity}(a), these channel
closures, together with the reduced vacancy production, account for
the decrease of $Y_{\mathrm{NEEC}}$ from
$2.67\times10^5$ to $1.12\times10^5$. In contrast,
$Y_{\mathrm{NEIES}}$ changes only weakly because NEIES does not
involve discrete electron-capture resonances and is affected mainly
through the modified configuration populations and electron
distribution. The corresponding NEEC fractions are shown in
Fig.~\ref{fig:S_B_sensitivity}(b) and listed in
Table~\ref{tab:S_B_sensitivity}. Over the full scan,
$F_{\mathrm{NEEC}}$ remains between $94.78\%$ and $97.99\%$.
NEEC therefore remains the predominant electron-induced excitation
channel even after the NEEC channels associated with the two largest
binding energies are energetically closed.

\begin{table}[h]
	\centering
	\caption{
		NEEC and NEIES yields, NEEC fractions, and numbers of
		configurations contributing to NEEC in the $6p$ binding-energy
		sensitivity scan.
	}
	\label{tab:S_B_sensitivity}
	\begin{tabular}{c c c c c}
		\hline\hline
		$\Delta B$ (eV) &
		$Y_{\mathrm{NEEC}}$ &
		$Y_{\mathrm{NEIES}}$ &
		$F_{\mathrm{NEEC}}$ (\%) &
		\shortstack{NEEC-contributing configurations} \\
		\hline
		$-0.50$ & $2.67\times10^{5}$ &
		$5.48\times10^{3}$ & $97.99$ & 6 \\
		$-0.25$ & $2.53\times10^{5}$ &
		$5.59\times10^{3}$ & $97.84$ & 6 \\
		$0$     & $2.40\times10^{5}$ &
		$5.71\times10^{3}$ & $97.68$ & 6 \\
		$+0.25$ & $1.66\times10^{5}$ &
		$5.83\times10^{3}$ & $96.61$ & 5 \\
		$+0.50$ & $1.62\times10^{5}$ &
		$5.95\times10^{3}$ & $96.45$ & 5 \\
		$+1.00$ & $1.12\times10^{5}$ &
		$6.19\times10^{3}$ & $94.78$ & 4 \\
		\hline\hline
	\end{tabular}
\end{table}

\bibliographystyle{apsrev4-2}
\bibliography{Sreference.bib}